\documentclass[a4paper]{jpconf}
\usepackage{graphicx}
\begin{document}
\title{Tau lepton physics at Belle}

\author{Kiyoshi Hayasaka (KMI, Nagoya U.)}

\address{Furo-cho, Chikusa-ku, Nagoya, Japan}

\ead{hayasaka@hepl.phys.nagoya-u.ac.jp}

\begin{abstract}
We report the recent results of a search for lepton-flavor-violating 
$\tau$ decays and a search for $CP$ violation 
in $\tau$ to $\nu K_S^0 \pi$ using a large data sample accumulated 
with the Belle detector at the KEKB asymmetric-energy 
$e^+e^-$ collider. 
The sensitivity to these modes is significantly 
improved compared to previous experiments.
\end{abstract}

\section{Introduction}
The success of the B-factory provides us
the world-largest data sample of $\tau^+\tau^-$-pairs,
that is useful for a search for New Physics (NP).
Here, we discuss new phenomena relating to a $\tau$ decays,
in particular, lepton-flavor-violating $\tau$ decays
and $CP$ violation in a $\tau$ decay using an around
1 ab${}^{-1}$ data sample
 accumulated with the Belle detector at the KEKB asymmetric-energy 
$e^+e^-$ collider, 

An observation of the charged-lepton-flavor violation (cLFV)
is a clear signature of NP since
the cLFV has a negligibly
small probability in the Standard Model (SM)
even taking into account neutrino oscillations
and many of models for NP naturally predict cLFV.
In particular, lepton-flavor-violating (LFV) $\tau$ decays
are allowed to have various kinds of final states and
some of them are expected to have a measurably large 
branching fraction, depending on the models.
Therefore, not only to find the NP phenomenon
but to distinguish the models, it is important
to search for the LFV $\tau$ decays.

Similarly to the LFV $\tau$ decays,
an observation of
 $CP$ violation (CPV) in a $\tau$ decay
is also a clear signature of NP.
Here, we focus on the CPV in 
a $\tau$ decay into $\nu K_S^0 \pi$
induced by a new scalar particle such as 
a charged Higgs in a multi-Higgs-doublet model.
This kind of CPV can be observed as a difference
in the $\tau^{\pm}$ decay angular distributions
by the interference effect between the scalar
and (axial-)vector components of the intermediate
state in the decay.

\section{\boldmath $\tau\rightarrow\ell M^0$ ($\ell=e,\mu$,
 $M^0=\pi^0,\eta,\eta^{\prime}, \rho^0, K^{\ast 0}, \bar{K}^{\ast
 0},\omega$ and $\phi$)}
Generally, $\tau\rightarrow\mu\gamma$ is the most promising among
various LFV $\tau$ decays in many models for NP. 
However, in some  models, such as non-universal-Higgs-mass 
model~\cite{Arganda:2008jj},
 $\tau\rightarrow\mu\eta$ can have the largest branching fraction.
 In addition, a magnitude relation among the branching fractions
for the LFV $\tau$ decays will help to distinguish the NP models.
Here, we show the results for the two-body $\tau$ decays
into a lepton and a neutral meson (pseudoscalar and vector).

\subsection{Analysis Method}

\begin{figure}[h]
\begin{center}
\includegraphics[width=0.5\textwidth]{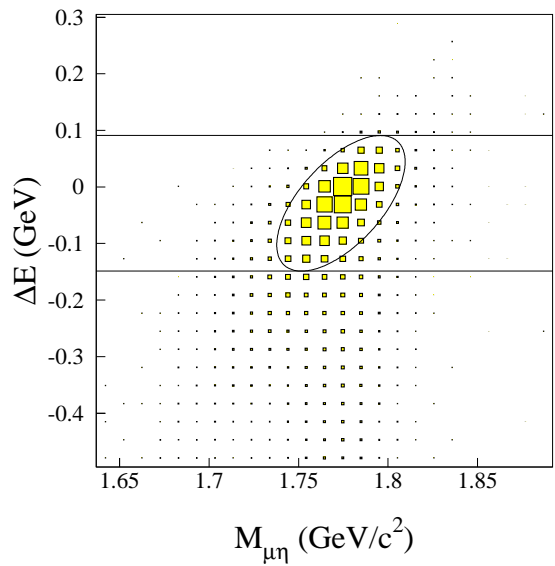}
\end{center}
 \caption{ Distribution for the $\tau\rightarrow\mu\eta$ signal
 events. An elliptical region is
defined as a signal region, that corresponds to $\pm3\sigma$ and
a region between two parallel lines excluding 
the signal region is used for the BG estimation as a sideband region,
where the resolution $\sigma$ is evaluated by a asymmetric-Gaussian fit
 to the signal MC distribution. }
\label{fig:signalMC}
\end{figure}

In the LFV $\tau$ analysis, in order to evaluate the number of
 signal events, two independent variables are defined,
that are
signal-reconstructed mass
and energy in the center-of-mass (CM) frame
from energies and momenta for the signal $\tau$ daughters.
In the $\tau\rightarrow\mu\eta$ case,
they are defined as
\begin{eqnarray}
 M_{\mu\eta}=\sqrt{E_{\mu\eta}^2-P_{\mu\eta}},\\
 \Delta E=E_{\mu\eta}^{\rm CM}-E_{\rm beam}^{\rm CM},
\end{eqnarray}
where $E_{\mu\eta}$ ($P_{\mu\eta}$) is a sum of the energies
(a magnitude of a vector sum of the momenta) for $\mu$ and $\eta$,
the superscript $\rm CM$ indicates that the variable is defined in the
CM frame and the $E_{\rm beam}^{\rm CM}$
means the initial beam energy in the CM frame.
Principally, $M_{\mu\eta}$ and $\Delta {E}$ are expected to be
$m_{\tau}$ ($\sim 1.78$ GeV/$c^2$) and 0 (GeV), respectively,
for signal events while $M_{\mu\eta}$ and $\Delta {E}$ will
smoothly vary without any special structure in the background (BG)
events.
The signal MC distribution of $\tau\rightarrow\mu\eta$
 on the $M_{\mu\eta}$ --
$\Delta{E}$ plane is shown in Fig.~\ref{fig:signalMC}.
Due to the resolution, the signal events are distributed
 around $M_{\mu\eta}\sim m_{\tau}$
and $\Delta E\sim0$ (GeV).
Taking into account the resolution, we set the elliptic signal region.
(See Fig.~\ref{fig:signalMC}.)
Since the resolution for the photon energy is not so good
as that for the momentum of the charged track,
the signal region of the mode having some photons
in the final state can be large.

To avoid any bias for our analyses, we perform the blind analysis,
i.e.,
we do not observe the data events in the signal region
before fixing the selection criteria and the evaluation
for the systematic uncertainties and the expected number
of the BG events.
Finally, we observe some signal events
or evaluate 
an upper limit of the number
of the signal event at 90\% confidence level (CL)
using the expected number of the
BG event and the number of the observed events
in the signal region by Feldman-Cousins
approach.~\cite{Feldman:1997qc}

\subsection{\boldmath $\tau\rightarrow\ell P^0$ ($\ell=e,\mu$,
  $P^0=\pi^0,\eta,\eta^{\prime}$)}

\begin{table}[t]
\footnotesize
\caption{Summary for the efficiency (Eff.),
the expected number of the BG events, ($N_{BG}^{\rm exp}$)
and the upper limit on the branching fraction (UL) for each mode,
where ``comb.'' means the combined result from subdecay modes.}
\begin{tabular}[t]{|l|c|c|c|l|c|c|c|}
\hline
 Mode ($\tau\rightarrow$) & Eff.(\%) &$N_{BG}^{\rm exp}$ & UL
 ($\times10^{-8}$)&
 Mode ($\tau\rightarrow$) & Eff.(\%) &$N_{BG}^{\rm exp}$ & UL
 ($\times10^{-8}$)\\
\hline
$\mu\eta(\rightarrow\gamma\gamma)$ &8.2&$0.63\pm0.37$&3.6&
$e\eta(\rightarrow\gamma\gamma)$ &7.0&$0.66\pm0.38$&8.2\\
\hline
$\mu\eta(\rightarrow\pi\pi\pi^0)$ &6.9&$0.23\pm0.23$&8.6&
$e\eta(\rightarrow\pi\pi\pi^0)$ &6.3&$0.69\pm0.40$&8.1\\
\hline\hline
$\mu\eta$(comb.) &&&2.3&
$e\eta$(comb.) &&&4.4\\
\hline\hline
$\mu\eta'(\rightarrow\pi\pi\eta)$ &8.1&$0.00^{+0.16}_{-0.00}$
&10.0&
$e\eta'(\rightarrow\pi\pi\eta)$ &7.3&$0.63\pm0.45$&9.4\\
\hline
$\mu\eta'(\rightarrow\gamma\rho^0)$ &6.2&$0.59\pm0.41$&6.6&
$e\eta'(\rightarrow\gamma\rho^0)$ &7.5&$0.29\pm0.29$&6.8\\
\hline\hline
$\mu\eta'$(comb.) &&&3.8&
$e\eta'$(comb.) &&&3.6\\
\hline\hline
$\mu\pi^0$ &4.2&$0.64\pm0.32$&2.7&
$e\pi^0$ &4.7&$0.89\pm0.40$&2.2\\
\hline
\end{tabular}
\label{table:lp0}
\end{table}
We perform a new search for the $\tau$ decay into a lepton ($e$ or
$\mu$) and a neutral pseudoscalar ($\pi^0$, $\eta$ or $\eta^{\prime}$)
with a 901 fb$^{-1}$ data sample, that corresponds to
$8.2 \times 10^8$ $\tau^+\tau^-$-pairs.
A neutral pion is reconstructed from 2 photons
while an $\eta$ ($\eta^{\prime}$)
is reconstructed from $\gamma\gamma$ ($\rho^0\gamma$)
as well as $\pi^+\pi^-\pi^0$ ($\pi^+\pi^-\eta$)
to increase the detection efficiency.
When the neutral pseudoscalars are reconstructed,
their four-momentum is evaluated by a mass-constrained fit
to obtain a better resolution for the signal region.
Because we have modified the selection criteria applied
to the previous analysis, we obtain an about $1.5$ times
better detection efficiency while a similar background 
level is kept. 
Here, the number of the BG events in the signal region
is evaluated by an interpolation from the number of the 
observed events in the sideband region, assuming
a flat distribution for the BG events in the signal
and sideband region.

\begin{figure}[b]
\includegraphics[width=\textwidth]{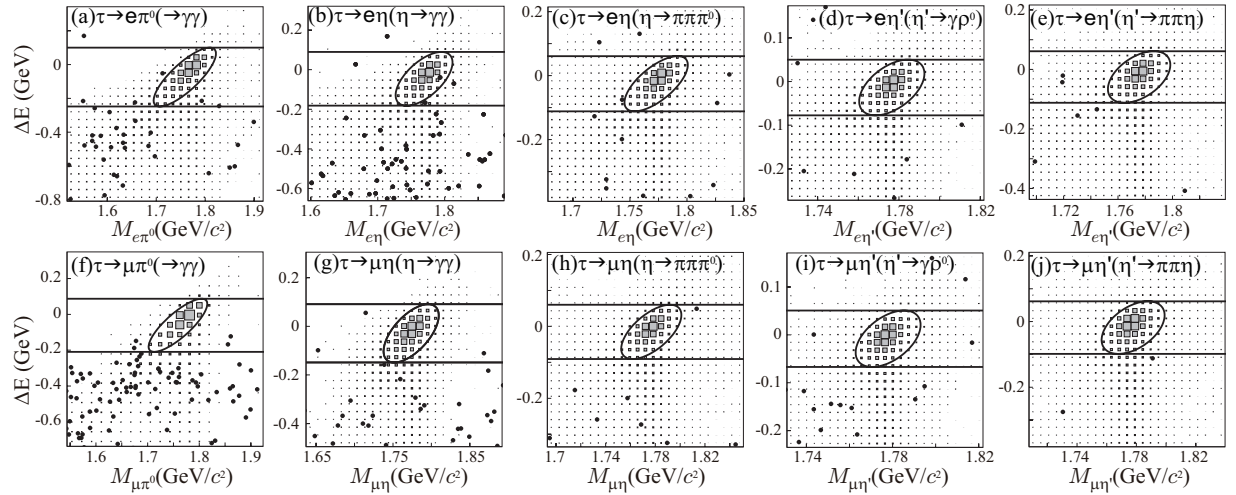}
\caption{\small Resulting 2D plots for 
$\tau\rightarrow e \pi^0$~(a),
$e \eta(\rightarrow\gamma\gamma)$~(b),
$e \eta(\rightarrow\pi\pi\pi^0)$~(c),
$e \eta^\prime(\rightarrow\gamma\rho^0(\rightarrow\pi\pi))$~(d),
$e \eta^\prime(\rightarrow\pi\pi\eta(\rightarrow\gamma\gamma))$~(e),
$\mu \pi^0$~(f),
$\mu \eta(\rightarrow\gamma\gamma)$~(g),
$\mu \eta(\rightarrow\pi\pi\pi^0)$~(h),
$\mu \eta^\prime(\rightarrow\gamma\rho^0(\rightarrow\pi\pi))$~(i)
and
$\mu \eta^\prime(\rightarrow\pi\pi\eta(\rightarrow\gamma\gamma))$~(j),
on the $M_{\ell P^0}$ -- $\Delta E$ plane. Here,
black dots (shaded boxes) express the data (signal MC),
the region bounded by two lines is defined
as a $3\sigma$ band for the BG estimation
and the elliptic region is the signal one which corresponds
to $3\sigma$
in each plot. One event is found in the signal region for
$\tau\rightarrow e \eta$
while no events appear in any other modes.}
\label{fig:lp0result}
\end{figure}

As a result, we observe one event
in the $\tau\rightarrow e\eta (\rightarrow\gamma\gamma)$ mode
while no events are found in other modes as shown 
in Fig.~\ref{fig:lp0result}. Since these results
are consistent with the background estimation,
we set upper limits 
on the following branching fractions,
as shown in Table~\ref{table:lp0}.
Consequently, we obtain a $(2.1-4.4)$ times more sensitive result
than previously.

\subsection{\boldmath $\tau\rightarrow\ell V^0$
($\ell=e,\mu$, 
$V^0=\rho^0,K^{\ast0},\bar{K}^{\ast0},\omega,\phi$)}
\begin{table}[t]
\footnotesize
\caption{Summary for the efficiency (Eff.),
the expected number of the BG events, ($N_{BG}^{\rm exp}$)
and the upper limit on the branching fraction (UL) for each mode.
}
\begin{tabular}[t]{|l|c|c|c|l|c|c|c|}
\hline
 Mode ($\tau\rightarrow$) & Eff.(\%) &$N_{BG}^{\rm exp}$ & UL
 ($\times10^{-8}$)&
 Mode ($\tau\rightarrow$) & Eff.(\%) &$N_{BG}^{\rm exp}$ & UL
 ($\times10^{-8}$)\\
\hline
$\mu\rho^0$ &7.1&$1.48\pm0.35$&1.2&
$e\rho^0$   &7.6&$0.29\pm0.15$&1.8\\
\hline
$\mu K^{\ast0}$ &3.4&$0.53\pm0.20$&7.2&
$e   K^{\ast0}$ &4.4&$0.39\pm0.14$&3.2\\
\hline
$\mu \bar{K}^{\ast0}$ &3.6&$0.45\pm0.17$ &7.0&
$e   \bar{K}^{\ast0}$ &4.4&$0.08\pm0.08$ &3.4\\
\hline
$\mu\omega$ &2.4&$0.72\pm0.18$&4.7&
$e  \omega$ &2.9&$0.30\pm0.14$&4.8\\
\hline
$\mu\phi$ &3.2&$0.06\pm0.06$&8.4&
$e\phi$   &4.2&$0.47\pm0.19$&3.1\\
\hline
\end{tabular}
\label{table:lV0}
\end{table}

\begin{figure}[b]
\includegraphics[width=\textwidth]{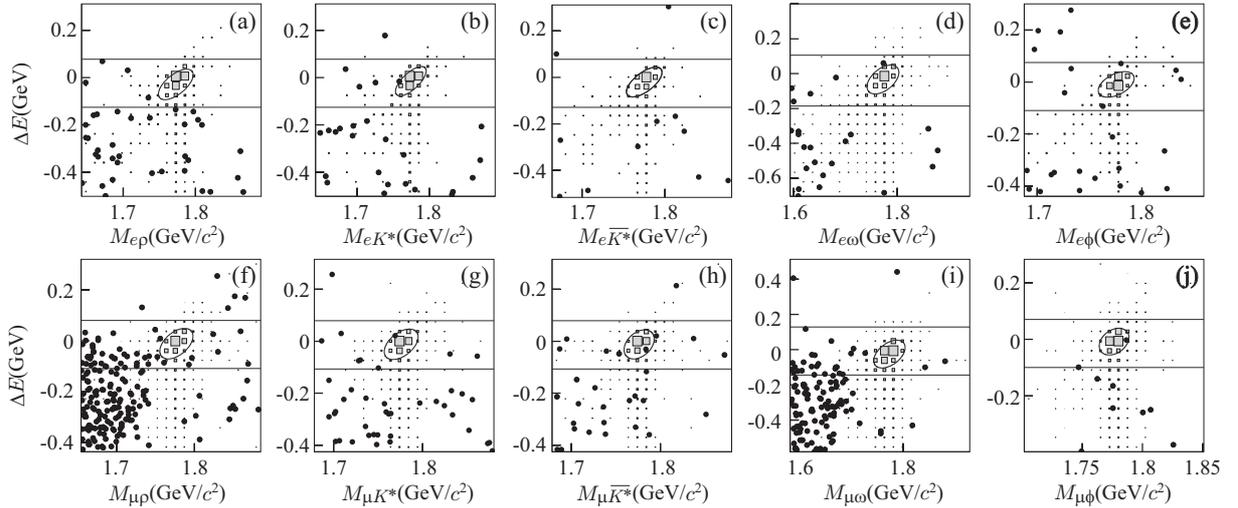}
\caption{\small Resulting 2D plots for 
$\tau\rightarrow e \rho^0$~(a),
$e K^{\ast0}$~(b),
$e \bar{K}^{\ast0}$~(c),
$e\omega$~(d),
$e\phi$~(e),
$\mu \rho^0$~(f),
$\mu K^{\ast0}$~(g),
$\mu \bar{K}^{\ast0}$~(h),
$\mu\omega$~(i) and
$\mu\phi$~(j)
on the $M_{\ell V^0}$ -- $\Delta E$ plane. Here,
black dots (shaded boxes) express the data (signal MC),
the region bounded by two lines is defined
as a $5\sigma$ band for the BG estimation
and the elliptic region is the signal one which corresponds
to $3\sigma$
in each plot. One event is found in the signal region for
$\tau\rightarrow\mu K^{\ast 0}$, $\mu \bar{K}^{\ast0}$ and $\mu\phi$
while no events appear in any other modes.}
\label{fig:lv0result}
\end{figure}

\begin{figure}[t]
\begin{center}
\includegraphics[width=\textwidth]{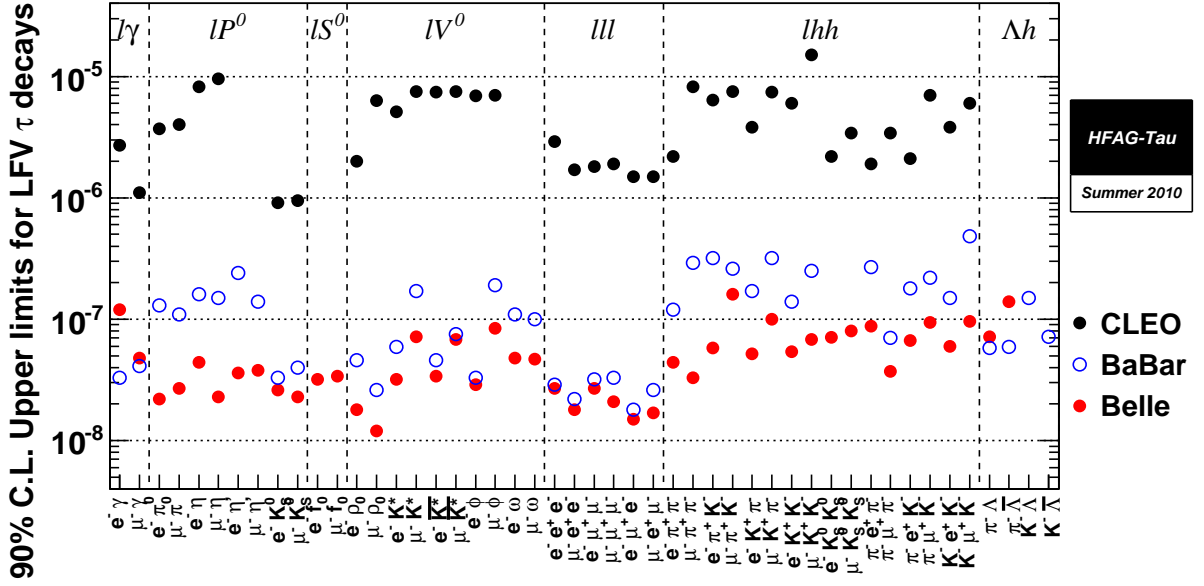}
\end{center}
\caption{Upper limits on the branching fractions for the LFV $\tau$
 decays. Here, red, blue and black circles express the results
obtained by Belle, BaBar and CLEO, respectively~\cite{HFAGtaulfv}.}
\label{fig:alllfv}
\end{figure}

Similarly to $\tau^-\rightarrow\ell^- P^0$, we update our results
of the search for $\tau^-\rightarrow\ell^- V^0$, 
where 
$\ell=e,\mu$, $V^0=\rho^0,K^{\ast0},\bar{K}^{\ast0},\omega,\phi$
with an 854 fb${}^{-1}$ data sample, that corresponds to
$7.8 \times 10^8$ $\tau^+\tau^-$-pairs~\cite{Miyazaki:2011xe}. Here,
$\rho^0,K^{\ast0},\bar{K}^{\ast0},\omega$ and $\phi$
are reconstructed from
$\pi^+\pi^-$, $\pi^-K^+$, $\pi^+K^-$,
$\pi^+\pi^-\pi^0$ and $K^+K^-$, respectively.
 By performing a detailed
background study, we obtain a 1.2 times better efficiency
in average with keeping similar level backgrounds.
For the modes including a $\mu$, BG events mainly come
from $\tau\tau$ events while some two-photon process
or radiative Bhabha process with a gamma conversion
becomes main background in the modes having an electron.
Finally, one event is found in the signal region for
$\tau^-\rightarrow\mu^- K^{\ast 0}$, $\mu^- \bar{K}^{\ast0}$ 
and $\mu^-\phi$
while no events appear in any other modes. 
(See Fig.~\ref{fig:lv0result}.)
They are consistent with the expected number of the backgrounds.
Consequently, we set the 90\% confidence level upper limits
on the branching fractions as shown in Table~\ref{table:lV0}.
At the present, the upper limit
 for $\tau\rightarrow\mu\rho^0$ is the most sensitive
among the all LFV $\tau$ decays.
(See Fig.~\ref{fig:alllfv}, where all of the current upper limits
 are shown.)

\section{CPV measurement with the $\tau$ decay into $\nu K_S^{0}\pi$}
In SM, CP in the lepton sector is conserved. However,
by introducing a new scalar field with a complex coupling constant,
CPV can be induced. A charged Higgs in the multi-Higgs-doublet 
model~\cite{mhdm}
is one of the candidates for such a scalar field. Here,
we discuss the CPV appearing in the $\tau\rightarrow\nu K_S^0 \pi$
decay as an interference between the SM and NP 
processes~\cite{Bischofberger:2011pw}.
\subsection{Analysis method}
A hadronic current into $K^0\pi$ can be written as
\begin{eqnarray}
 J_\mu&=&\langle K^0(q_1)\pi(q_2)|\bar{u}\gamma_\mu s | 0 \rangle
\nonumber \\
&=& (q_1-q_2)^\nu\left(g_{\mu\nu}-\frac{Q_\mu Q_\nu}{Q^2}\right)
F_V(Q^2)+Q_\mu F_S(Q^2),
\end{eqnarray}
where $u$ and $s$ are spinor fields for a $u$ and $s$ quark,
respectively,
$q_1$ and $q_2$ are 4-momenta for $K^0$ and $\pi$,
respectively,
$F_V$ $(F_S) $ is a vector (scalar) form factor 
and $Q^\mu=q_1^\mu+q_2^\mu$.
To make the current induce CPV,
instead of $F_S$, we introduce $\tilde{F}_S$, defined as
\begin{equation}
 \tilde{F}_S= F_S + \frac{\eta_S}{m_\tau} F_H,
\end{equation}
where $F_H$ is a form factor for the charged Higgs,
$\eta_S$ is a complex coupling constant~\cite{kandm}. 
The size of CPV is proportional to
an imaginary part of $\eta_S$.
From the current, in the hadronic rest frame ($\vec{Q}=\vec{0}$),
we can calculate the differential
partial width for $\tau\rightarrow\nu K_S^0 \pi$ as
\begin{eqnarray}
 \frac{d\Gamma(\tau\rightarrow\nu K_S^0 \pi)}{d\omega}
&=& \mbox{(CP even terms)} \nonumber \\
&& -4 \frac{m_\tau}{\sqrt{Q^2}}|\vec{q_1}| \Im(F_V F_H^{\ast}) \Im(\eta_S)
\cos\psi \cos\beta,
\end{eqnarray}
where $\omega=dQ^2d\cos\theta d\beta$,
 $\theta$ ($\beta$) is an opening angle between
the direction opposite to the one of the CM system
and the one of the hadronic system in the $\tau$ rest frame
(between the direction of $K_S^0$ and that for the CM system
in the hadronic rest frame),
and $\psi$ denotes the angle between the direction of the
CM frame and the direction of the $\tau$ as seen from 
the hadronic rest frame.

To extract the CP violating term, we define an asymmetry as
\begin{eqnarray}
 A^{CP}_{\psi\beta} &=& \frac{\int \cos \beta \cos \theta
\left(\frac{d\Gamma_{\tau^-}}{d\omega}
-\frac{d\Gamma_{\tau^+}}{d\omega}
\right) d\omega}{\frac12
\left(\frac{d\Gamma_{\tau^-}}{dQ^2}
+\frac{d\Gamma_{\tau^+}}{dQ^2}
\right)dQ^2
} \nonumber \\
&\simeq& \langle\cos\beta\cos\psi \rangle_{\tau^-}
- \langle\cos\beta\cos\psi \rangle_{\tau^+}.
\end{eqnarray}
Therefore, by measuring the second line,
we can evaluate  $A^{CP}_{\psi\beta}$ experimentally.
\subsection{Result and Discussion}
\begin{figure}[b]
 \begin{center}
\includegraphics[width=0.4\textwidth]{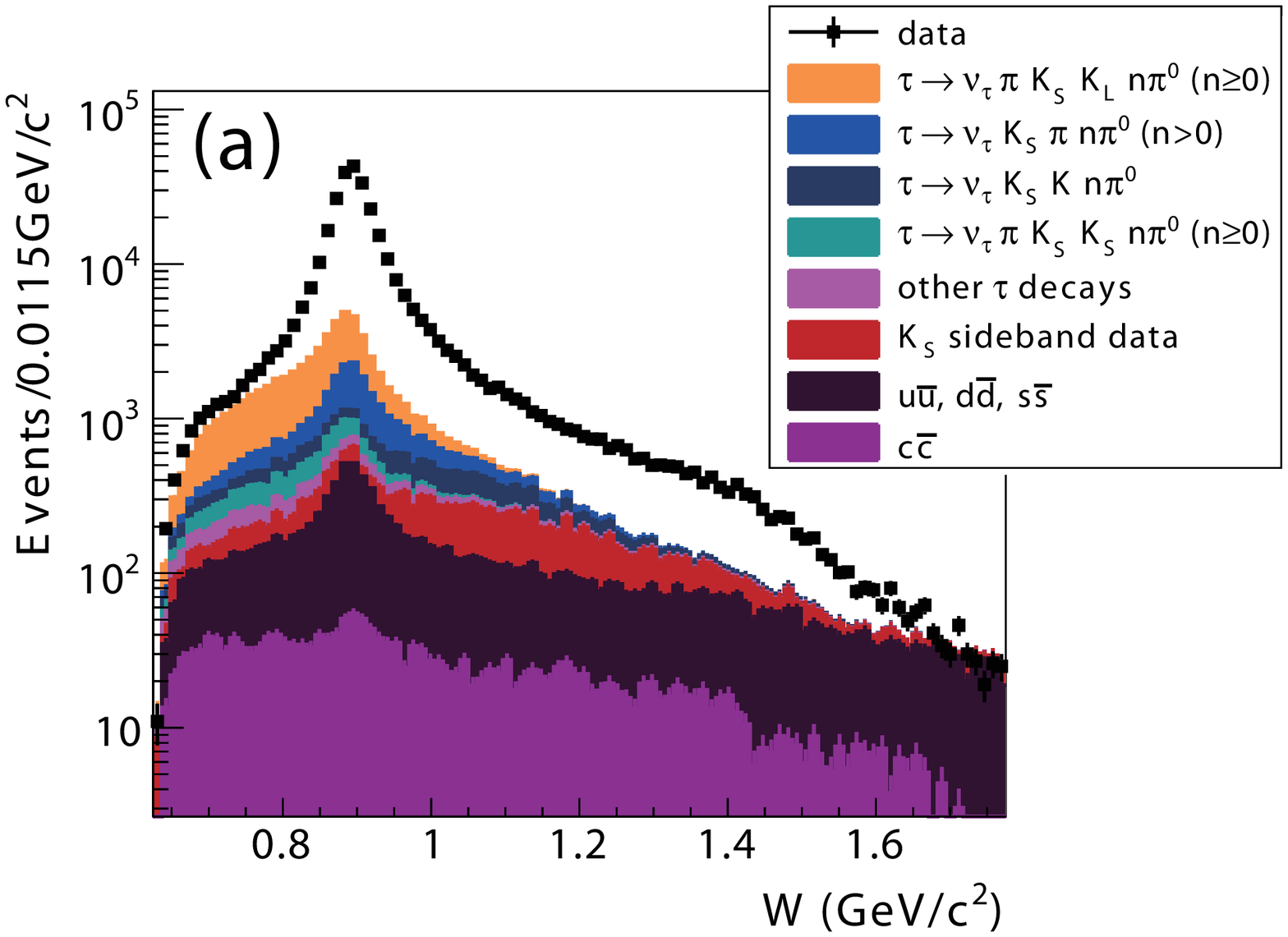}  
\includegraphics[width=0.28\textwidth]{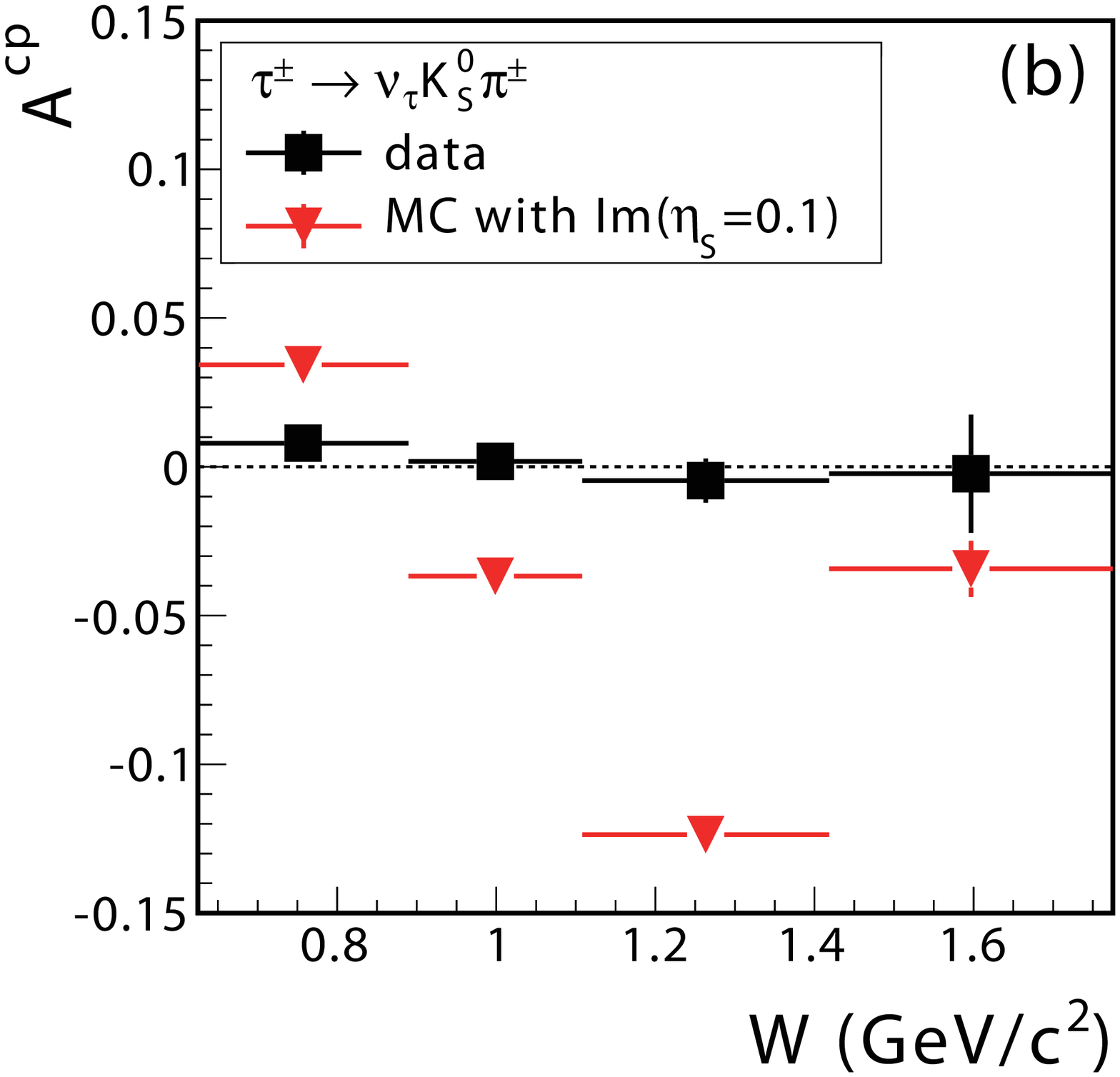}  
\includegraphics[width=0.28\textwidth]{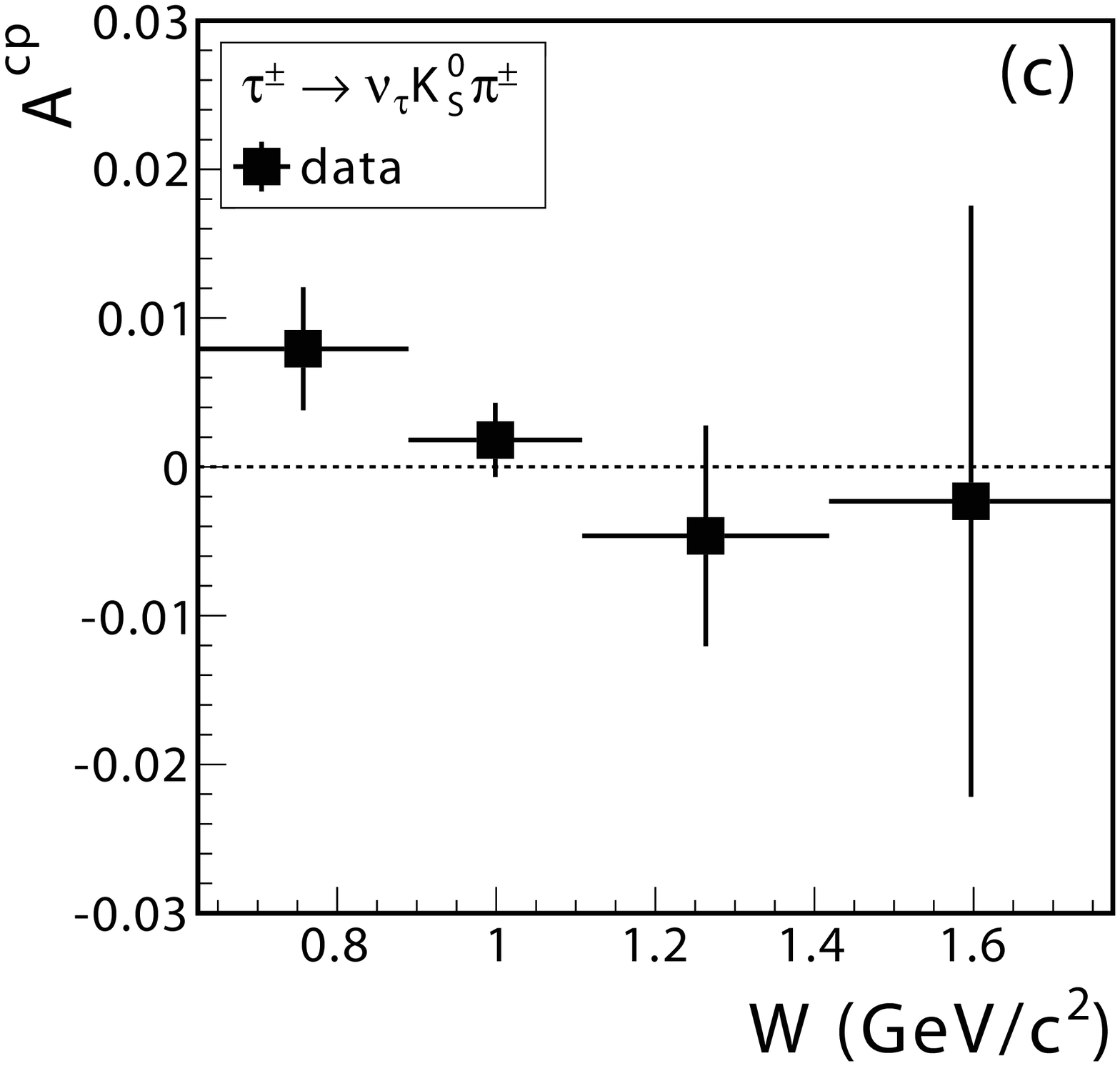}  
\caption{(a) Distribution of $W$ for the $\tau\rightarrow\nu K_S^0\pi$
candidate,
(b) plot for $A^{CP}_{\beta\psi}$ depending on $W$ and
(c) zoom of (b).}
\label{fig:CPV}
 \end{center}
\end{figure}

With a 699 fb${}^{-1}$ data sample,
we obtain around $3\times10^5$ $\tau\rightarrow\nu K_S^0\pi$ candidates,
where it turns out that 23\% of them are BG events by MC study.
(See Fig.~\ref{fig:CPV}(a).)
Using them, $A^{CP}_{\psi\beta}$ is evaluated depending on
$\sqrt{Q^2}(\equiv W)$. (See Table~\ref{table:CPV} and
Figs.~\ref{fig:CPV}(b)
and (c).)  In the final result,
$\gamma-Z$ interference and $\pi^{\pm}$ detection asymmetry
effects are corrected for and the background subtraction is
also performed.
\begin{table}[h]
 \begin{center}
\caption{ CP asymmetry in bins of the hadronic mass $W$.
The first and second errors correspond to statistical and systematic ones,
respectively.}
  \begin{tabular}{|c||cccc|}
\hline
   & \multicolumn{4}{|c|}{W (GeV/$c^2$)}\\
\cline{2-5}
& 0.625--0.890
& 0.890--1.110
& 1.110--1.420
& 1.420--1.775 \\
\hline\hline
$A^{CP}_{\beta\psi}$ 
&$7.9\pm3.9\pm2.8$
&$1.8\pm2.1\pm1.4$
&$-4.6\pm7.2\pm1.7$
&$-2.3\pm19.1\pm5.5$\\
\hline
  \end{tabular}
 \end{center}
\label{table:CPV}
\end{table}

Consequently, no CPV is found. From this result,
we can evaluate the 90\% CL upper limit for $|\Im(\eta_S)|$
varying a constant-strong-interaction-phase difference
between $F_V$ and $F_S$:
 \begin{equation}
|\Im(\eta_S)|<(0.012 - 0.026) .
 \end{equation}
This result is almost one order of magnitude more restrictive
than that previously obtained by CLEO~\cite{cleocpv}, i.e., 
$|\Im(\eta_S)|<0.19$.

According to a theoretical prediction for $\Im(\eta_S)$ in the
multi-Higgs-doublet model, $\eta_S$ is written as
\begin{equation}
 \eta_S \simeq \frac{m_\tau m_s}{M^2_{H^{\pm}}} X^* Z,
\end{equation}
where $m_s$ is an $s$-quark mass,
$M_{H^{\pm}}$ is the lightest charged Higgs mass,
and $X$ and $Z$ are complex constants describing
the coupling of the Higgs to the
$u$ and $s$ quarks and the $\tau$ and $\nu$, respectively.
Using the limit $\Im(\eta_S)<0.026$, we can obtain an inequality:
\begin{equation}
 |\Im(XZ^*)| < 0.15\frac{M^2_{H^{\pm}}}{1 \mbox{GeV${}^2$/$c^4$}}.
\end{equation}

\end{document}